\documentclass[aps,prc,showpacs,tightenlines,twocolumn,floatfix]{revtex4}

\usepackage{braket}
\usepackage{graphicx}
\usepackage{amsmath}
\usepackage{amssymb}
\usepackage{verbatim}

\def\vec#1{\boldsymbol{#1}}
\newcommand{\nuc}[2]{\hbox{$^{#1}$#2}}

\newcommand{\hr}{\hat{\vec{r}}}

\newcommand{\cg}[1]{(#1)}

\newcommand{\pwt}[2]{\phi_{\ell_{#1}j_{#1}}^{\lambda_{#1}}({#2})}

\newlength{\figwidth}

\begin{document}
\setlength{\figwidth}{0.9\columnwidth}

\title{Two-nucleon correlation effects in knockout reactions from \nuc{12}C}

\author{E.~C. Simpson}
    \affiliation{Department of Physics, Faculty of Engineering and
      Physical Sciences, University of Surrey, Guildford,
      Surrey GU2 7XH, United Kingdom}
\author{J.~A. Tostevin}
    \affiliation{Department of Physics, Faculty of Engineering and
      Physical Sciences, University of Surrey, Guildford,
      Surrey GU2 7XH, United Kingdom}

\date{\today}
\begin{abstract}
Reactions that involve the direct and sudden removal of a pair of
like or unlike nucleons from a fast projectile beam by a light
target nucleus are considered. Specifically, we study the three
two-nucleon removal channels from $^{12}${C} that populate final
states in the $^{10}${Be}, $^{10}${B} and $^{10}${C} reaction
residues. The calculated two-nucleon removal cross sections and the
residue momentum distributions are compared with available high
energy data at 250, 1050 and 2010 MeV per nucleon, data that are
inclusive with respect to the bound final-states of the residues. The
measured $np$-removal cross sections only are significantly greater
than the values calculated, suggesting that the reaction mechanism
observes enhanced $np$ spatial correlations compared to those
present in the shell-model wave functions.
\end{abstract}
\maketitle

\section{Introduction}

There is an extensive literature on the development and the use of
high-energy electron-induced knockout of both one or two nucleons
from the ground states of stable nuclei as probes of nucleonic
correlations and of pair correlations. Relevant and recent reviews 
include references \cite{ee1,ee2,rev2,BD}, which cite the essential 
experimental developments and data sets: see also the citations in
\cite{SSM08}. Quite recently, new conclusions have been drawn
regarding the observation of enhanced short-ranged correlations
(SRC) between nucleon-nucleon pairs from $(e,e'pN)$ experiments on a
carbon target \cite{SSM08}. Specifically, measurements that selected
high momentum transfer and large missing momentum events from the
few-body final-state phase-space found evidence that $np$ pairs were
more than an order of magnitude more prevalent than like-nucleon
pairs in this selection; deduced from the corresponding $(e,e'np)$
and $(e,e'2p)$ yields. The probabilities of these high relative
momentum two-nucleon components have been computed, using the 
microscopic ground-state wave functions from {\em ab-initio}
variational Monte Carlo calculations, for systems with masses $A
\leq 8$ \cite{Bob1,Bob2} and, using a linked cluster expansion,
for $A \geq 12$ \cite{Alv}. These computations involve the 
wave functions of the nucleons over the entire volume of the nuclei
concerned. These theoretical studies are strongly suggestive that
the tensor force, in the spin $S=1$, isospin $T=0$, $np$ channel,
plays the major role in generating these enhanced $np$-spatial
correlations and their associated high-relative-momenta signatures
in the two-nucleon density distribution. In such $(e,e'pN)$ 
measurements, the specific final-state phase space
selection (of high relative momentum back-to-back two-nucleon
events) provides the leverage and the amplification of the
short-ranged pair-correlation sensitivity.

The behavior of the wave functions of single nucleons and of pairs
of nucleons (in a mass $A+2$ projectile) can also be probed if the
nucleons are removed (suddenly) in fast collisions with a light
target nucleus. Within this strong-interaction probe case, the
sensitivity is now to the nuclear wave functions at and near the
nuclear surface. Such processes are also referred to as one- and
two-nucleon knockout (or removal) reactions. These direct reaction
mechanisms, combined with $\gamma$-decay spectroscopy, to determine
the final state of the reaction residues, and their momentum
distributions, are being actively exploited as a spectroscopic tool
to study the evolution of nucleonic single-particle structure near
the two (often very displaced) Fermi surfaces of exotic nuclei. They
are proving to be robust techniques, see e.g.
\cite{HaT03,BHS02,TBB04}.

Unlike the final-state-exclusive measurements in electron induced
reactions, observables in such nuclear-induced fast nucleon removal
reactions are inclusive with respect to the final states of the
removed nucleon(s) and of the fate of the struck light target
nucleus. The cross sections are thus relatively large. Measurements
usually consist of the total removal reaction yield, often the
momenta of the fast, forward travelling projectile-like residues,
and sometimes the differential yields to the ground and bound
excited states of the mass $A+1$ or mass $A$ residual nuclei; the
latter obtained by $\gamma$-ray spectroscopy. However, in such
collisions between the projectile and a light composite target
nucleus, such as beryllium or carbon, the reaction is {\em
geometrically} very selective \cite{STB09b}, and the removal of two nucleons will 
be enhanced if nucleon pairs have a strong spatial correlation (and
localization) in the projectile ground state. An interesting
question therefore is whether like and unlike two-nucleon removal
under such conditions also exhibits any evidence of enhanced $np$
over $nn$ and $pp$ spatial correlations on a longer length scale
than implied by the (SRC) observations of the electron knockout
data? Here, the spatial (geometrical) selectivity of the reaction
mechanism would provide the leverage and probe of the presence of
spatially localized pairs and evidence of an enhanced $np$
correlation.

Experimental data are available in the form of high-energy, primary
beam measurements of the inclusive cross sections to the bound states
of the residues after
$np$, $nn$ and $pp$ removal \cite{KLC88,LGH75}. We will show that
these data reveal a significant enhancement of unlike-pair yields,
$\sigma_{-np}$, relative to those, $\sigma_{-nn}$ and
$\sigma_{-pp}$, for like-nucleon pairs, the enhancement being
significantly greater than would be expected (trivially) from the
numbers of such pair combinations available.

The removal of two (well-bound) nucleons of the deficient species
from asymmetric nuclei has been shown to proceed as a direct
reaction \cite{BBC03}. A description of the reaction making use of
configuration-mixed shell-model wave functions and elastic and
inelastic breakup contributions to the removal cross section has
been shown to reproduce experiment \cite{ToB06}. Such calculations
have also recently been extended to describe the reaction-residue
momentum distributions \cite{STB09a,STB09b}, where it was shown that
the shapes and widths of the residue momentum distributions are
indicative of both the total angular momentum ($I$) and the total
orbital angular momentum components $(\vec{L}= \vec{\ell}_1 +
\vec{\ell}_2$) of the removed nucleon pair \cite{SiT10}.

Here we will exploit the eikonal reaction model in the isospin
formalism \cite{TPB04,ToB06,STB09b} for the removal of like ($T=1$)
and unlike ($T=0,1$) nucleon pairs. We also discuss the possible
role of indirect population of the residue final states of interest,
i.e. by single-nucleon knockout paths that populate particle-unbound
states of the intermediate mass $A$+1 system. Such indirect paths
severely limit the number of systems that, in the absence of an
empirical means to distinguish between direct and indirect knockout
events, may be studied quantitatively with unlike pair knockout.
These indirect contributions to final state yields are expected, in
general, to be large in $np$ removal from asymmetric systems
\cite{SiT09} where one or other of the nucleon thresholds will be at
relatively low excitation energy.

Attractive $np$-removal test cases thus suggest symmetric light
nuclei, e.g. \nuc{12}{C} or \nuc{16}{O}. For such less-massive cases
one must however consider core-recoil effects in the reaction and
the analogous center-of-mass corrections to the (fixed center)
shell-model two-nucleon amplitudes.  Both are expected to affect, in
detail, the absolute magnitudes of the cross sections. These are
discussed here. Regardless of these two (relatively small) effects,
comparisons of the relative like- and unlike-pair knockout yields
may be made.

In Section \ref{sec:carbon} we outline the specific features of the
reactions using $^{12}$C projectiles. The necessary formalism has
been presented elsewhere and will only be outlined in Section
\ref{sec:formalism}, exploiting the notation used in previous work.
The calculated results for \nuc{12}{C} are discussed in Section
\ref{sec:results} and a summary made in Section \ref{sec:summary}.

\section{Carbon induced reactions \label{sec:carbon}}

Our current expectation is that a quantitative discussion of
pair-correlation effects can only be made when the residue final
states are populated predominantly via a single-step direct
reaction. It is thus vital to select examples that minimize the
indirect (evaporative) contributions to the reaction yield: and that
we assume cannot (at present) be distinguished from the direct
reactions of interest. We must therefore consider systems with large
(and symmetric) nucleon separation thresholds, such that the
single-nucleon removal strength to particle-unbound states will be
weak.  For the same reason, we require the projectile nucleus to be
relatively light to minimize population of the (bound) final residue
states via the removal of deeply bound (non-valence) nucleons.
Ideally, any experiment would also measure single-nucleon and
like-pair removal in addition to the unlike pair removal, in order
to verify shell- and reaction-model predictions for the distribution
of single-nucleon spectroscopic strength and for the direct $T=1$
pair knockout. The above considerations severely constrain the
potential candidates, the best examples being \nuc{12}{C} and
\nuc{16}{O}. The former of these will be considered in detail here.
The choice of an $N=Z$ projectile poses the additional
(experimental) complication of distinguishing the mass $A$ residue
from the incident beam, the residue and projectile having identical
mass-to-charge ratios.

Consideration of two-nucleon knockout from $^{12}$C is valuable for
two reasons. First, its shell model description and that of the
residual nuclei $^{10}$C, $^{10}$Be and $^{10}$B are extensively
studied and so establish a valuable point of reference. Second, the
existing experimental cross sections for two-nucleon knockout from
$^{12}$C \cite{LGH75,KLC88} are accurate to $\approx$ 10\% and were
taken at high energies, where the eikonal model used here is at its
most reliable. The experimental data were obtained using reactions
of a carbon beam on a carbon target at 250, 1050 and 2100 MeV per
nucleon incident energies. These show an expected enhancement of the
$^{10}$B production cross sections ($np$ removal) over those for the
$nn$ (to $^{10}$C) and $pp$ (to $^{10}$Be) removal reactions (see
Table \ref{12C}). The significance of this observed enhancement is
quantified here.

\begin{figure}[tb]
\includegraphics[width=\figwidth]{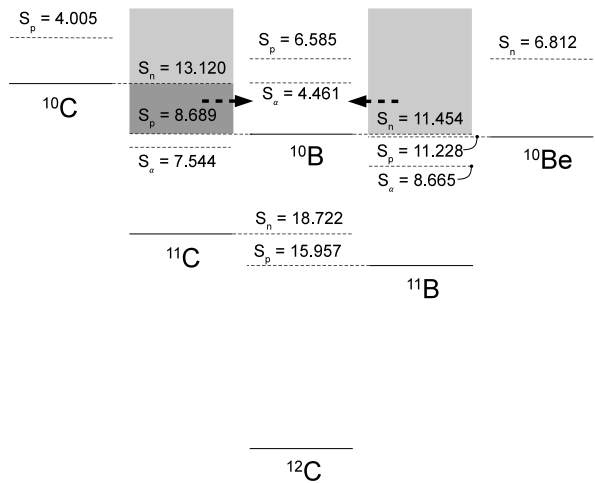}
\caption{One- and two-nucleon and $\alpha$-particle separation
thresholds relevant to nucleon-removal reactions from \nuc{12}{C}.
The \nuc{12}{C} two-nucleon separation energies are $S_{2p}=27.184$,
$S_{2n}=31.184$ and $S_{np}=27.412$ MeV. Proton evaporation from the
single-neutron removal residue $^{11}$C would be expected to be the
largest indirect pathway, but the shell-model calculations suggest
very little spectroscopic strength to states above 8.689 MeV in
\nuc{11}{C}. \label{fig:c12_seps}}
\end{figure}

In addition to two-nucleon removal, single-nucleon removal cross
sections were also measured at high energy (and previously studied
theoretically, see Ref. \cite{BHS02}). These verify the expectation,
from the shell-model, that the majority of single-particle removal
strength is exhausted in transitions to final states lying below the
nucleon separation thresholds of $^{11}$C and $^{11}$B. This point
will be developed further in later Sections. The relevant one- and
two-nucleon (and $\alpha$-particle) separation thresholds are
illustrated in Fig. \ref{fig:c12_seps}.

\begin{figure}[tb]
\includegraphics[width=1.02\figwidth]{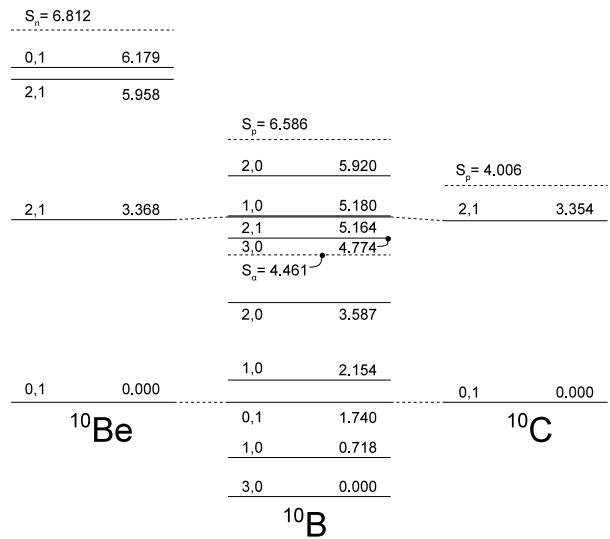}
\caption{States of the mass $A=10$ residues populated in the
two-nucleon knockout. The spin and isospin labels ($J_f$,$T_f$) are
indicated. All states included are of positive parity.  Levels
assumed to be part of the isospin multiplet are connected by
dashed lines. The lowest particle thresholds are also
indicated. States above the $\alpha$-particle threshold in
\nuc{10}{B} are expected to decay via $\alpha$ emission, with the
exception of the 5.164 MeV, $T=1$, $J^\pi=2^+$ state which has an
84\% $\gamma$-decay branch. \label{fig:levels}}
\end{figure}

The simplest of the three two-nucleon removal cases is two-neutron
($nn$) removal, leading to the $^{10}$C residual nucleus. This has
only two bound final states below the first proton threshold, the
0$^+$ ground state and a 2$^+$ excited state at 3.354 MeV
\cite{TKG04}. The level scheme of $^{10}$Be, the residual nucleus in
the $pp$ removal case, is only slightly more complex. We must
consider population of the 0$^+$ ground state, the 2$^+$ states at
3.368 and 5.958 MeV, and a second 0$^+$ state at 6.179 MeV, all
below the neutron threshold of 6.812 MeV. The most complicated final
state is that for $^{10}$B, the $np$ knockout residue. The $p$-shell
shell-model calculations used include even-parity states up to a
maximum of spin 3. For \nuc{10}{B} we have shown states up to the
proton separation threshold, however the low $\alpha$-particle
separation threshold means that the $T=0$ states at 4.774 (3$^+$),
5.180 (1$^+$) and 5.920 MeV (2$^+$) are reported to decay (with
branching ratios of near 100\%) by $\alpha$-emission. We will show
the cross sections for population of these states but will assume
that they do not contribute to the calculated \nuc{10}{B} yield. The
$T=1$, 2$^+$(5.184 MeV) state is reported to have a 16\% $\alpha$
emission branch. These \nuc{10}{B} states are illustrated in Fig.
\ref{fig:levels}. We note that the \nuc{10}{B} spectrum also
contains several negative parity states which are not expected to be
populated by the nucleon removal reaction mechanism.

Direct, unlike-pair-removal cross sections are expected to be larger
than those for like-pairs. Assuming the simplest $p-$shell $\pi[0p_{3/2}]^4
\nu[0p_{3/2}]^4$ structure for \nuc{12}{C}, the pn-format TNAs for pair
removal to a residue final state with spin $J_f$ are given by the
appropriate coefficients of fractional parentage, with value
$\sqrt{2J_f+1}$. The cross sections to a given final state, being
proportional to the square of these TNA, suggest unlike:like
two-nucleon inclusive cross section ratios $\sigma_{-np}/
\sigma_{-2N}= 16/6 \approx 2.7$ if all strength leads to bound final
states. The same result is obtained from the counting of available
valence-nucleon pairs.

The primary motivation for the present study is that this simple
combinatorics expectation is not corroborated by the available data;
the ratio $\sigma_{np}/\sigma_{pp}$ ($\sigma_{np}/\sigma_{nn}$) from
the experiments at 250, 1050 and 2100 MeV/nucleon are 8.1 (8.9), 5.3
(6.3) and 6.0 (8.5), respectively \cite{LGH75}, the observed
enhancement of the $np$ removal cross section being significantly
larger than expected.

Clearly our simple estimate, from the number of available pairs of
each type, assumes that all of the removal strength leads to bound
configurations in the residue of interest, which is not the case,
and part of the enhancement noted above may be attributable to this
(channel-dependent) fraction of events leading to particle-unbound
states. Here we seek to address this question quantitatively, by use
of a direct reaction model for two-nucleon removal. Specifically,
our aims are to quantify: (i) whether this enhancement is accounted
for by those two-nucleon correlations that are included in the
truncated, $p$-space shell model calculation, and (ii) to make a
simple estimate of the contributions one might reasonably expect
from indirect reaction pathways in this model picture.

\section{Formalism \label{sec:formalism}}

\subsection{Two-nucleon overlap \label{sec:overlap}}

The formalism used is based on that developed in Refs.
\cite{ToB06,STB09b}. Isospin-format two-nucleon amplitudes (TNAs)
will be used and are expected to offer a good description of the
light symmetric systems considered here. Thus, in the unlike pair
removal case, we assume a common set of nucleon orbital wave
functions, to be discussed below.

We evaluate the cross sections for transitions from the projectile
initial (ground) state $i$, with spin $(J_i, M_i)$, to particular
residue final states $f$. The residue is assumed to be a spectator
in the sudden reaction description and its state is not coupled to
the reaction dynamics. The direct reaction will then probe the
two-nucleon overlap. We denote the $A$-body final states by
$\Phi^{(F)}(A)$, where the label $F\equiv(f,M_f)$ includes the
angular momentum projection $M_f$. The two-nucleon wave function
(two-nucleon overlap) of the removed nucleons 1 and 2 is written as
$\Psi_i^{(F)}(1,2)$, where
\begin{align}
\Psi_i^{(F)} \equiv& \Psi_{J_i M_i T_i \tau_i}^{(F)}(1,2)\nonumber
\\ \equiv & \braket{\Phi^{(F)}(A)
|\Psi_i (A,1,2)} \nonumber\\ =&\sum_{I\mu{T}\alpha}C_{\alpha}^{IT}
\cg{I{\mu}J_fM_f|J_iM_i} \nonumber\\
&\cg{T{\tau}T_f\tau_f|T_i\tau_i} \;
[\,\overline{\psi_{\beta_1}(1)\otimes \psi_{\beta_2}(2)}\,]_{I
\mu}^{T\tau}\ . \label{eqn:overlap}
\end{align}
Here the set of available active two-nucleon configurations, with
counter $\alpha$, consists of particular pairs of single-particle
orbitals [$\beta_1,\,\beta_2$], where the index $\beta=(n \ell j)$
denotes the state's spherical quantum numbers. The $C_{\alpha}^{IT}$
are the two-nucleon amplitudes (TNAs) that express the parentage of
the residue final state, when coupled to a particular two-nucleon
configuration $\alpha$, in the projectile ground state. Here, as
elsewhere, these are taken from a truncated-basis shell model
calculation.

\begin{figure}[tb]
\includegraphics[width=\figwidth]{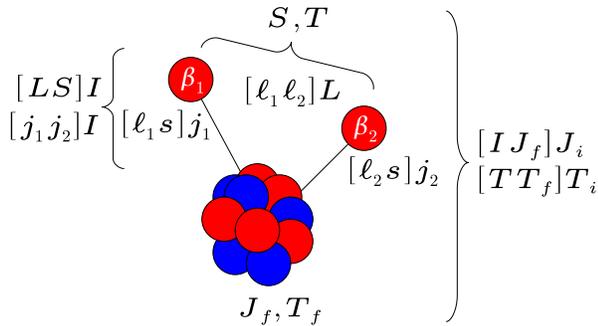}
\caption{(Color online) Diagram of the angular momentum and isospin
couplings, and the coupling orders assumed. Specifically, the
$LS$-couplings used in this paper are presented.
\label{fig:coupling}}
\end{figure}

Expressed in $LS$-coupling, the antisymmetrized two-nucleon wave
function of Eq. (\ref{eqn:overlap}) is
\begin{widetext}
\begin{align}
[\,\overline{\psi_{\beta_1}(1)\otimes\psi_{\beta_2}(2)}\,]_{I\mu}^{T
\tau} =& D_{\alpha}\hat{j}_1\hat{j}_2 \sum_{\substack{L\Lambda{S\Sigma}\\
\lambda_1\lambda_2}}\cg{\ell_1\lambda_1\ell_2\lambda_2|L\Lambda}
\cg{L\Lambda{S}\Sigma|I\mu} \,  \hat{L}\hat{S} \, \chi_{S\Sigma}
(1,2) \chi_{T\tau} (1,2)\nonumber\\ \times & [\pwt{1}{1}
\pwt{2}{2}-(-)^{S+T} \pwt{1}{2}\pwt{2}{1}]\left\{
\begin{array}{ccc} \ell_1 & s & j_1 \\  \ell_2 & s & j_2 \\
L & S & I \\ \end{array} \right\}\ , \label{eqn:anti_wf_ls}
\end{align}
\end{widetext}
with $D_{\alpha}=1/\sqrt{2(1+\delta_{\beta_1\beta_2})}$. The angular
momentum and isospin couplings used are summarized in Fig.
\ref{fig:coupling}. The nucleon-wave functions $\pwt{}{i}$ are
\begin{align}
\pwt{}{i}=u_{\beta}(r_i)Y_{\ell\lambda}(\hr_i)\ . \label{eqn:wf_ls}
\end{align}

We note that no explicit account is taken of additional (short,
medium, long range) correlations other than those contained within
the shell-model description, and that enter Eq. (\ref{eqn:overlap}).
These are seen to arise from: (a) antisymmetry and angular momentum
coupling of the nucleon pair, and (b) the (shell-model) two-nucleon
overlap, via the weights and phases of the contributing TNAs. If
there are significant additional strong-interaction generated pair
correlations, missing from our description, then we might expect
empirical cross sections to deviate significantly (and be enhanced)
relative to the shell-model-correlated model used here.

\subsection{Centre-of-mass corrections to the TNA}

Spectroscopic factors calculated within the fixed-center shell-model
basis require a center-of-mass correction factor to be applied
\cite{DiF74}. Essentially, the shell-model two-nucleon overlaps are
calculated relative to the centre of mass of all $A+2$ nucleons and
not relative to the remaining $A$ nucleons. Such corrections were
discussed by Pinkston \cite{Pin76,Pin77,PiI79} (see also Ref.
\cite{BGP85}). In previous, predominantly $sd$-shell applications
such corrections were expected to be small, but for the examples
discussed here they may be more significant, particularly when one
has an ambition to compare absolute cross sections with experiment.

In the simplest (harmonic oscillator) limit, the requirement is to
multiply the shell-model TNAs by a factor $[(A+2)/A]^{N/2}$, with
$N$ the number of oscillator quanta of the orbitals of the active
nucleons. In the present case, of \nuc{12}{C}($-$2N), this would
enhance all cross sections by $\approx 20\%$, independent of the
reaction channel and the residue. Thus, conclusions regarding the
{\em relative strengths} of the different two-nucleon removal
channels will be unaffected and we do not apply these corrections.
Clearly, a more complete understanding will be necessary for a
precise consideration of the absolute cross sections.

\subsection{Reaction dynamics and approximations}

We exploit eikonal reaction dynamics with the required nucleon- and
residue-target $S$-matrices calculated in the optical limit of
Glauber theory \cite{Gla59,ATT96}. These elastic $S$-matrices are
thus calculated assuming that the residue and nucleons travel on
straight line paths through the interaction field of the target. The
residue-target $S$-matrices are calculated by double-folding their
densities with an effective nucleon-nucleon (NN) interaction. The
absorptive nature of these interactions naturally localizes the
reaction to the projectile surface \cite{STB09b}. A detailed derivation of the
two-nucleon removal cross section is presented elsewhere
\cite{ToB06}, as is the formalism for their momentum distributions
\cite{STB09b,SiT10}.  These are not reproduced here.

The key approximations are as follows. The removal of nucleons is
sudden, i.e. their co-ordinates are assumed to vary slowly compared
to the timescale of the reaction, and can be assumed frozen during
the interaction. In previous work we have also (reasonably) made the
no-recoil (heavy residue) approximation in the diffraction-stripping
terms of the removal cross section \cite{ToB06}; i.e. we have
assumed coincidence of the residue and projectile centers of mass
and impact parameters, $b_c \approx b$.  These core recoil effects
will certainly introduce corrections for the lighter residues in
this work, however, as will be quantified later, at the high
projectile energies considered here these diffractive-stripping
terms contribute a significantly smaller fraction of the cross
section than the two-nucleon stripping terms. Thus, recoil will have
a relatively minor effect on the calculated removal cross sections
and the discussion of this paper. This effect will be fully
quantified in future work.

\section{Nucleon knockout from \nuc{12}{C} \label{sec:results}}

\subsection{Methodology \label{sec:method}}

The experimental information available consists of final state
inclusive cross sections to bound states of the mass $A=11,10$
residues and some associated residue momentum distribution
information. We are able to identify each known positive parity
final state ($A=10$) and negative parity final state ($A=11$) with a
well-defined shell model state. The shell-model calculations were
performed with {\sc oxbash} \cite{BEG04}. We will consider the
results when using the {\sc wbp} interaction \cite{wbp}, and also
compare these with the {\sc pjt} interaction \cite{pjt} outcomes, in
the $p-$shell model space. The former interaction was used in
previous work on nucleon removal from \nuc{12}{C} \cite{BHS02}. The
shell-model spectroscopic factors $C^2 S$ and the two-nucleon
spectroscopic amplitudes, the $C_\alpha^{IT}$, as enter Eq.
(\ref{eqn:overlap}), were calculated for each residue final state.
To discuss possible indirect processes we must consider the
population of highly-excited, unbound states in the single-nucleon
removal cases.

The $p-$shell model space allows removal of nucleons from the active
$p_{3/2}$ and $p_{1/2}$ single-particle orbitals. The radial wave
functions used to describe these nucleonic states are calculated in
a Woods-Saxon potential well. The radius and diffuseness parameters
were fixed to be $r_0=1.310$ fm and $a=0.55$ fm, as were used in the
single-nucleon knockout analysis of Ref.\ \cite{BHS02}. In addition,
we include a spin-orbit potential of depth $V_{so}=6$ MeV. The
ground-state to ground-state $nn$, $pp$ and $np$ separation
energies, $S_{2N}$, from \nuc{12}{C} are 31.841 MeV, 27.184 MeV and
27.41 MeV, respectively. The nucleon bound state wave functions for
each reaction and transition, with final-state excitation energy
$E_f$, were calculated using effective nucleon separation energies
of
\begin{align}
B_f = S_N + E_f\ ,
\end{align}
for single nucleon removal and
\begin{align}
B_f = (S_{2N} + E_f)/2
\end{align}
for each nucleon in two-nucleon removal. In the $np$-removal case we
assumed a common set of radial wave functions for the neutrons and
protons. These were calculated as above but assuming each nucleon
carried a charge of 0.5$e$.

The required nucleon- and residue-target $S$-matrices were
calculated using the optical limit of Glauber theory \cite{Gla59} -
i.e. by double folding each constituent particle and target density
with an effective nucleon-nucleon (NN) interaction. The essential
input parameters, given the dominance of the stripping mechanism,
are the free $nn$ ($\equiv pp$) and $np$ cross sections, taken from
the parametrization of Ref. \cite{ChG90}, and the
residue and target densities. The former are given in Table
\ref{param}. The effective NN interaction was assumed to be
zero-range, represented by a $\delta$-function, and the
real:imaginary forward scattering amplitude ratios $\alpha_{nn}$ and
$\alpha_{np}$ were obtained from a polynomial fit to the values tabulated in Ref.
\cite{Ray79}. These are listed in Table \ref{param} for the three
energies of interest. The point-nucleon density distributions of the
target and the residues were assumed to have Gaussian shapes, with
root mean squared (rms) radii consistent with Glauber-model analyses
of the measured interaction cross sections \cite{OST01}.  For simplicity,
we assume a single matter rms radius for each mass nucleus;
for $^{12}$C target and $A=11$ and $A=10$ residues, the mass radii
were 2.32, 2.11 and 2.30 fm respectively.  The latter value is that quoted
for \nuc{10}{Be} in Ref. \cite{OST01}.

\begin{table}[tbp]
\caption{Nucleon-nucleon effective interaction parameters used for
the calculation of the nucleon- and residue-target $S$-matrices.
\label{param}}
\begin{ruledtabular}
\begin{tabular}{ccccc}
Energy (MeV/u)&$\sigma_{nn}$ (mb)&$\sigma_{np}$ (mb)&$\alpha_{nn}$&
$\alpha_{np}$\\ \hline
250&22.3&36.6&~~0.94&~~0.49\\
1050&49.6&41.6&$-$0.08&$-$0.46\\
2100&63.9&46.5&$-$0.20&$-$0.47\\
\end{tabular}
\end{ruledtabular}
\end{table}

\subsection{Single-nucleon removal cross sections}

We first briefly review the single-nucleon removal results, as were
previously discussed in Ref. \cite{BHS02}. Of particular interest
here is the proportion of spectroscopic strength exhausted below
particle separation thresholds, and hence the shell-model prediction
of single-particle strength lying above the nucleon separation
thresholds. For the {\sc wbp} interaction 3.93 units of
spectroscopic strength are associated with states below these
thresholds with the remaining 0.07 fragmented over many states above
10 MeV in excitation. Calculations using the {\sc pjt} interaction
distribute the single-particle spectroscopic strength similarly,
with 3.97 units associated with states below the mass $A=11$ nucleon
thresholds.

The results for single-nucleon removal are very similar to those
presented previously \cite{BHS02}. The small differences can be
traced to our inclusion of a spin-orbit term in the nucleon bound
states potential. The detailed decomposition of the single-particle
cross sections for each final state into their stripping and
diffraction components, for the beam energy of 2100 MeV/nucleon, are
shown in Table \ref{tbl:c12_1n}. A discussion of the single-nucleon
shell-model suppression factors can be found in Ref. \cite{BHS02}.
Here we simply note that the theoretical cross section overestimates
the experiment by approximately a factor of two, consistent with the
results from electron-induced proton knockout \cite{BD}.

\begin{table}[tb]
\begin{ruledtabular}
\caption{Single-nucleon knockout cross sections (mb) for the
\nuc{12}{C} projectile incident on a carbon target at 2100 MeV per
nucleon. The single-nucleon removal cross sections presented,
$\sigma_{-1N}$, include both the spectroscopic $C^2S$ (from the {\sc
wbp} interaction) and center of mass correction factors (here
$12/11$). The results are very similar to those of Ref.
\cite{BHS02}.  The inclusion here of the nucleon bound states
spin-orbit potential accounts for the very minor differences
observed. \label{tbl:c12_1n}}
\begin{tabular}{cccccc}
Residue & $J^\pi_f$ & $\sigma_{str}$ & $\sigma_{dif}$ & $C^2S$ &
$\sigma_{-1N}$  \\ \hline
\nuc{11}{C} & 3/2$^-$ & 20.50 & 2.17 & 3.16 & 78.18 \\
& 1/2$^-$ & 18.45 & 1.81 & 0.58 & 12.82 \\
& 3/2$^-$ & 18.74 & 1.81 & 0.19 & 4.26 \\
\cline{2-6}
&&&& sum &  95.36 \\
&&&& exp. &  46.50 $\pm$2.30 \\
\hline
\nuc{11}{B} & 3/2$^-$ & 21.11 & 2.30 & 3.16 & 80.70 \\
& 1/2$^-$ & 18.86 & 1.90 & 0.58 & 13.14 \\
& 3/2$^-$ & 19.09 & 1.88 & 0.19 & 4.35\\
\cline{2-6}
&&&& sum & 98.81\\
&&&& exp. & 53.80 $\pm$2.70 \\
\end{tabular}
\end{ruledtabular}
\end{table}

With regards to possible indirect contributions to $np$ knockout, we
can estimate an upper bound assuming that the remaining shell-model
$C^2S$ strength (of 0.07) from both proton and neutron channels will
populate the $np$ residue indirectly. Doing so we obtain an indirect
contribution of $\approx 3$ mb. It is however unlikely that this
entire residual $p$-shell strength would lead to the \nuc{10}{B}
residue. The removal of deeply-bound $0s_{1/2}$ nucleons may also
lead to the \nuc{10}{B} residue, though the single-particle cross
section will be significantly smaller than those presented in Table
\ref{tbl:c12_1n} due to the stronger binding. We return to this
indirect contribution discussion in the following Section.

\subsection{Two-nucleon removal cross sections}

The stripping mechanism component of the two-like-nucleon removal
cross sections were previously discussed in Ref. \cite{TBP04}. Here
we extend these results to consider the stripping-diffraction
mechanism contributions and consideration of unlike-pair knockout to
confront the available experimental data quantitatively. As was
discussed earlier, disregarding the configuration mixing inherent in
the shell-model two-nucleon overlap, we might expect the unlike pair
knockout yields to be enhanced by approximately a factor of 16/6
over those for like nucleon knockout, from the counting of available
$[p_{3/2}]^2$ pairs. As was also pointed out, this zeroth-order
estimate assumes all of the two-nucleon strength will fall below the
first particle-emission thresholds. In practice the number of bound
states and the fraction of the removal strength leading to bound
states is different for the three residues. The lowest relevant
particle separation thresholds are as follows: \nuc{10}{C},
$S_p=4.006$ MeV; \nuc{10}{Be} $S_n=6.586$ MeV; \nuc{10}{B},
$S_{\alpha}=4.461$ MeV. Both the number of pair combinations and the
distribution of two-particle removal strength should be reasonably
accounted for when using the shell model two-nucleon amplitudes.

Results for the inclusive cross sections for two-nucleon removal
from \nuc{12}{C} to the three residues, at beam energies of 250,
1050 and 2100 MeV per nucleon, are given in Table \ref{12C}. The
final-state-exclusive like and unlike two-nucleon removal cross
sections, and their decomposition with the contributing reaction
mechanisms (i.e. stripping, $\sigma_{str}$, diffraction-stripping,
$\sigma_{ds}$, and estimated diffraction, $\sigma_{dif}$) are shown
in Table \ref{tbl:c12_2n} for the 2100 MeV per nucleon case and when
using the {\sc wbp} interaction TNAs. Table \ref{12C} shows the
calculated theoretical like-pair removal cross sections $\sigma_{th
}$, to \nuc{10}{C} and \nuc{10}{Be}, are in reasonable agreement
with the experimental data, $\sigma_{exp}$, of Refs.\
\cite{LGH75,KLC88}.

\begin{table*}[tbp]
\caption{Theoretical and experimental cross sections for two nucleon
knockout from $^{12}$C, for projectile energies of 250, 1050 and
2100 MeV per nucleon. All cross sections are in mb. The TNAs used
were calculated using the {\sc wbp} interaction. \label{12C}}
\begin{ruledtabular}
\begin{tabular}{c|ccc|ccc|ccc}
Energy&\multicolumn{3}{c|}{\nuc{10}{Be}} &
\multicolumn{3}{c|}{\nuc{10}{C}}&
\multicolumn{3}{c}{\nuc{10}{B}} \\
MeV/u&$\sigma_{th}$ &$\sigma_{exp}$ & $\sigma_{exp}/\sigma_{th}$ &
$\sigma_{th}$ &$\sigma_{exp}$ & $\sigma_{exp}/\sigma_{th}$ &
$\sigma_{th}$ &$\sigma_{exp}$ & $\sigma_{exp}/\sigma_{th}$ \\
\hline 250 \cite{KLC88} & 7.48 &5.88$\pm$9.70& 0.79$\pm$1.30
&5.80&5.33$\pm$0.81& 0.92$\pm$0.14
& 21.57 &47.50$\pm$2.42 & 2.20$\pm$0.11\\
1050 \cite{LGH75}& 6.62 &5.30$\pm$0.30& 0.80$\pm$0.05 &5.13
&4.44$\pm$0.24&0.87$\pm$0.05
& 19.27 &27.90$\pm$2.20 & 1.45$\pm$0.11\\
2100 \cite{LGH75} &6.52&5.81$\pm$0.29& 0.89$\pm$0.04 &5.04
&4.11$\pm$0.22&  0.82$\pm$0.04
&19.02&35.10$\pm$3.40 & 1.84$\pm$0.18\\
\end{tabular}
\end{ruledtabular}
\end{table*}

From Table \ref{tbl:c12_2n} we note that, at this projectile energy,
the two-nucleon stripping (absorption) term, $\sigma_{str}$,
accounts for $\approx$70\% of the calculated cross section. The
cross sections for those $T=1$ states common to all three residues,
namely the first 0$^+$ and 2$^+$ states, are also essentially equal,
the minor differences in the calculations arising from the small
differences in the separation energies for each system.

For $np$ removal to \nuc{10}{B}, the cross sections are shown for
the nine $p$-shell shell-model final states below the first nucleon
threshold. However, the first 2$^+$, $T=1$ state is known to decay
by $\alpha$-emission with a branch of $I_\alpha=16\%$. For this
state, this branching has been accounted for in the $\sigma_{-2N}$
value presented. The cross sections are also shown for the three
highest energy $T=0$ states, but these states are reported to decay
(with $I_\alpha =100\%$) by $\alpha$-emission. These cross sections
are not included in the summed $^{10}$B yield. These exclusive
\nuc{10}{B} cross sections will be discussed further in Section
\ref{sec:mom} and in Figure \ref{fig:10c}, in connection with their
associated momentum distributions. We note here that if one uses the
{\sc pjt} shell-model interaction TNAs then the summed $^{10}$B
yield is 18.73 mb, compared to the 19.02 mb shown in Table
\ref{tbl:c12_2n}. The calculated inclusive cross sections are thus
robustly determined within the $p$-shell model space calculations.

\begin{table}[htb]
\caption{Like and unlike two-nucleon removal cross sections (in mb)
for a \nuc{12}{C} projectile incident on a carbon target at 2100 MeV
per nucleon. The excitation energies, $E_f$, of each final state are
shown in Fig. \protect\ref{fig:levels}. The TNAs used were
calculated using the {\sc wbp} interaction. The sums show the
accumulated cross sections that lead to the ground state and the
$\gamma$-decaying bound excited states of the mass $A=10$ residues.
\label{tbl:c12_2n}}
\begin{ruledtabular}
\begin{tabular}{ccccccc}
Residue & $J^\pi_f$ & $T$ & $\sigma_{str}$ & $\sigma_{ds}$ &
$\sigma_{dif}$ & $\sigma_{-2N}$  \\ \hline
\nuc{10}{C} & 0$^+$ & 1 & 1.59 & 0.64 & 0.06 & 2.30 \\
& 2$^+$ & 1 & 1.96 & 0.71 & 0.06 & 2.74 \\
\cline{2-7}
& & &  & & sum & 5.04 \\
& && & & exp. & 4.11$\pm$0.22 \\
\hline
\nuc{10}{Be} & 0$^+$ & 1 & 1.65 & 0.68 & 0.07 & 2.40 \\
& 2$^+$& 1 & 2.02 & 0.74 & 0.07 & 2.83 \\
& 2$^+$& 1 & 0.88 & 0.32 & 0.03 & 1.23 \\
& 0$^+$& 1 & 0.04 & 0.01 & 0.00 & 0.06 \\
\cline{2-7}
& & & & & sum & 6.52 \\
&& & & & exp. &  5.81$\pm$0.29 \\
\hline
\nuc{10}{B}
& 3$^+$ & 0& 5.11 & 2.00 & 0.20 & 7.30   \\
& 1$^+$ & 0& 2.47 & 1.01 & 0.10 & 3.58 \\
& 0$^+$ & 1& 1.62 & 0.66 & 0.07 & 2.35 \\
& 1$^+$ & 0& 1.81 & 0.69 & 0.07 & 2.57 \\
& 2$^+$ & 0& 0.63 & 0.24 & 0.02 & 0.89 \\
& 3$^{+a}$ & 0& 1.14 & 0.43 & 0.04 & 1.62 \\
& 2$^{+b}$ & 1& 1.99 & 0.72 & 0.07 & 2.33 \\
& 1$^{+a}$ & 0& 0.30 & 0.10 & 0.01 & 0.41 \\
& 2$^{+a}$ & 0& 0.75 & 0.28 & 0.03 & 1.05 \\
\cline{2-7}
& & & & & sum & 19.02 \\
&& & & & exp. & 35.10$\pm$3.40 \\
\end{tabular}
\end{ruledtabular}
\footnotetext[1]{states decay by $\alpha$-emission with a 100\%
branching ratio.} \footnotetext[2]{state decays by $\alpha$ emission
with a 16\% $\alpha$-branch.}
\end{table}

As for the present like-nucleon cases, previous analyses of like
two-nucleon removal data for asymmetric $sd-$shell nuclei have
indicated a suppression of the measured values compared to the
theoretical model being used. Previously this factor was of
order $R_s (2N) = \sigma_{exp}/\sigma_{th} \approx 0.5$. The present
higher energy, like-nucleon calculations suggest a larger value of
$R_s$, see Table \ref{12C}. As was noted earlier, the present
calculations do not include center-of-mass (recoil) corrections to
the two-nucleon amplitudes or the reaction dynamics. The latter will
introduce minor corrections to the diffraction-stripping term only,
that contributes to the cross sections at the 30\% level. Since the
magnitude of these effects is not quantified here, we do not draw
detailed conclusions regarding the absolute cross sections and
$R_s(2N)$ for \nuc{12}{C}, although the required corrections will
not be large. We show the ratio $\sigma_{exp} /\sigma_{th}$ in Table
\ref{12C} to compare the results for the relative strengths of the
different two-nucleon removal channels. All three channels will be
affected similarly by the corrections mentioned above.

We note that in the full calculations there is an enhancement in the
theoretical $np$ removal cross sections for all three incident
energies, already in excess of the $16/6\approx2.7$ ratio from the counting of
pairs. The calculated $np:pp$ cross section ratios, $\sigma_{np}
/\sigma_{pp}\approx 2.9$, are marginally smaller than those for the
$np:nn$ case, $\approx 3.8$, largely reflecting the different
fractions of spectroscopic strength to bound states in the two
like-nucleon cases - there being only two bound states in
\nuc{10}{C}.

Since the magnitudes of the $pp$ and $nn$ removal cross sections are
reasonably described, our expectation is that the cross sections to
the states of the $T=1$ isospin multiplet in \nuc{10}{B} are similarly
well determined. However, an independent measurement of these cross
sections would provide a useful verification of the direct nature of
the reaction. Thus, within the direct reaction model used, we must
attribute the cross section deficit to the calculated yields of the
$T=0$ final states. We also note that, whilst the energy dependence
of the like-pair removal cross sections is reasonably reproduced by
the theoretical calculations, there is some discrepancy in the
unlike-pair removal channel - with the result that the degree of
underestimation varies with energy. The theoretical values for both
the like- and unlike-pair removal decrease moderately with
increasing projectile energy. This said, there remain considerable
variations from the experimental $np$ knockout cross sections.

\subsection{Momentum distributions \label{sec:mom}}

In addition to the fragment production cross section measurements
of Ref. \cite{LGH75}, residue momentum distributions were also
measured \cite{GLH75}. Except for the case of two-proton removal
using a \nuc{9}{Be} target, the experimental data sets have been
published only as the widths of Gaussian fits to the experimental
data. Here we denote these as $\Delta_G$. Further, the published
width parameters were the averages of data measured for several
targets (Be, CH$_2$, C, Al, Cu, Ag and Pb), though it was observed
that, within the accuracy of the experiment, there was no dependence
on the target mass above a 5\% level. The references above do not
discuss the thicknesses of the targets used, or of the momentum
resolution of the incident beam, both of which could broaden the
measured widths. We assume, since the data are for high-energy
(stable) beams, that this experimental broadening of the width is
negligible.

Following, and to enable comparisons with the results of Ref.
\cite{GLH75}, we fit our calculated (projectile rest frame) {\em
inclusive} residue momentum distributions with Gaussian functions.
The two-nucleon stripping and diffractive-stripping mechanisms are
expected to yield very similar residue momentum distributions \cite{STB09b},
and here we present calculations for pure stripping events, scaled
to match the mechanism-inclusive cross section for each final state.
In general these calculated final-state-inclusive residue momentum
distributions are very close to Gaussian shapes. This is not the
case however for the individual (exclusive) residue final states,
particularly those of higher spin which have a broader and flatter
distribution near the central value, $\kappa_c=0$ MeV/c. The results
are collected and compared in Table \ref{tbl:gauss} and are,
encouragingly, in good agreement.

\begin{table}[htbp]
\caption{Projectile rest frame residue momentum distribution widths
$\Delta_G$ (for reactions at 2100 MeV per nucleon) obtained by
fitting Gaussian profiles to the experimental \cite{GLH75} and
theoretical (this work) inclusive distributions to bound final
states. The TNAs used were calculated using the {\sc wbp}
interaction. The data of Ref.\ \cite{GLH75} are the average widths
from data for several targets (see text). The calculated values are
for a carbon target only. \label{tbl:gauss}}
\begin{ruledtabular}
\begin{tabular}{ccc}
Residue & $\Delta_G^{exp}$ & $\Delta_G^{th}$  \\
\hline
\nuc{11}{B} & 106$\pm4$  & 99  \\
\nuc{11}{C} & 103$\pm4$  & 100 \\
\hline
\nuc{10}{Be}& 129$\pm$4  & 127  \\
\nuc{10}{B}&  134$\pm$3  & 132 \\
\nuc{10}{C}&  121$\pm$6  & 120 \\
\end{tabular}
\end{ruledtabular}
\end{table}

\begin{figure}
\includegraphics[width=\figwidth]{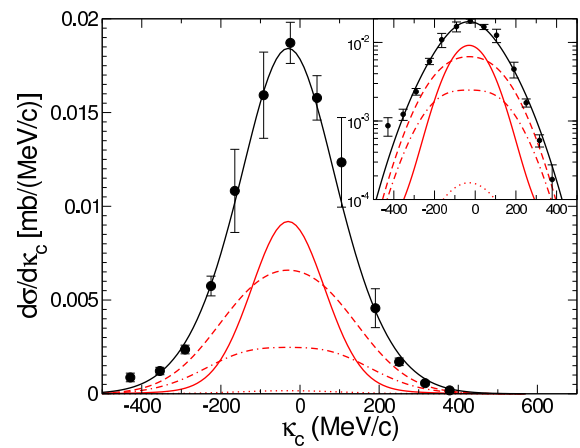}
\caption{(Color online) Comparison of measured and calculated 
\nuc{10}{Be} projectile frame residue momentum distributions.
The red curves show the \nuc{10}{Be} final-state-exclusive results
for the ground state (solid line), $2_1^+$ (dashed line), 
$2_2^+$ (dot-dashed line) and $0_2^+$ (dotted line). The inclusive
cross section, their sum, is shown by the black solid line. The data 
is for a \nuc{9}{Be} target whereas the calculations assume a \nuc{12}{C}
target. The calculations have been offset by $-$30 MeV/c and the
have been scaled to match the experimental two-proton removal cross
section on the \nuc{9}{Be} target, of 5.97 mb. The TNAs used were
calculated using the {\sc wbp} interaction.\label{fig:greiner}}
\end{figure}

The calculated residue momentum distributions for two-proton removal
are compared to the data in Fig. \ref{fig:greiner}, the data having
been scanned from Figure 1 of Ref. \cite{GLH75}. The data provided
are for a beryllium target while the calculations assume a carbon
target. Any differences in the momentum distributions are expected
to be very small for these two light targets. Both the exclusive
contributions from the four \nuc{10}{Be} final states involved and
their sum are shown. The figure shows clearly that the inclusive
Gaussian-like distribution is formed of exclusive cross sections of
very different widths for the different final states. Thus, residue
momentum distributions for these individual final states would be of
considerable value in validating, or revealing deficiencies in the
details of the shell-model wave functions and TNAs used here,
including our description of the reaction as a direct single-step
process.

\begin{figure}
\includegraphics[width=0.95\figwidth]{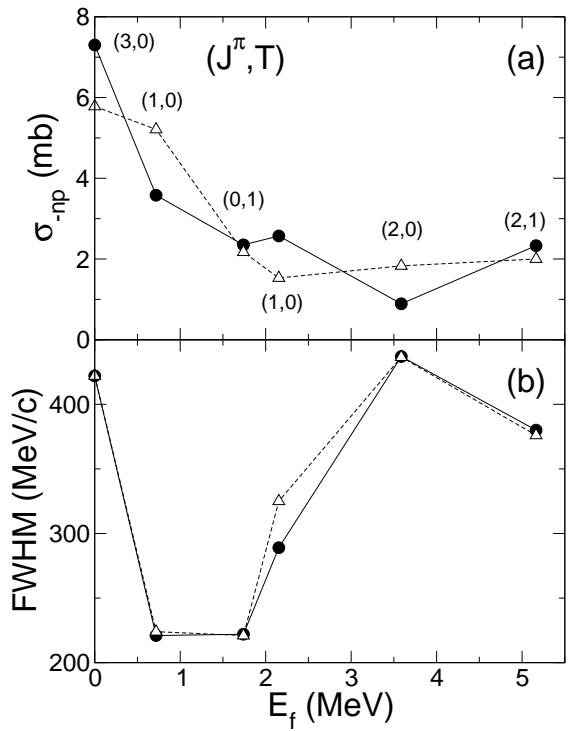}
\caption{Calculated exclusive cross sections (a) and full
width at half maximum (FWHM) widths of the momentum distributions
(b) for the six $\gamma$-decaying final states of the
\nuc{10}{B} residues, for reactions at 2100 MeV per nucleon. The
TNAs used were obtained using the {\sc wbp} interaction (solid lines
and solid points) and the {\sc pjt} interaction (dashed lines and
open points). The significant variations predicted: (i) for the
widths of the momentum distributions to the different final states,
and (ii) in the sensitivity of the \nuc{10}{B}, $T=0$ final state
yields to the effective interaction used, are evident.}
\label{fig:10c}
\end{figure}

Details of these calculated full width at half maximum (FWHM) widths
in the case of $np$-removal at 2100 MeV per nucleon, for the six
$\gamma$-decaying final states of \nuc{10}{B}, are shown in Fig.
\ref{fig:10c}. Results from both the {\sc wbp} (solid lines and
filled points) and {\sc pjt} (dashed lines and open points)
shell-model interaction TNA are shown. The widths from the two
interactions are remarkably similar for all transitions, indicative
that the $LS$-contributions to each final state are very similar 
\cite{SiT10}. The momentum distributions for the first and second
$T=0$, 1$^+$ states of \nuc{10}{B} were also presented in Figure 3
of Ref. \cite{SiT10}, calculated from the {\sc wbp} interaction. The
corresponding exclusive cross sections (upper panel of the figure)
also show very similar trends for the two shell-model interactions.
They differ in detail, however, particularly with regard the ratio
of the cross sections to the ground $(3^+, T=0)$ and first excited
$(1^+, T=0)$ states. We note also that the two interactions predict
very similar cross sections in the two $T=1$ states, the differences
being more evident in the $T=0$ states yields. Once again, this
suggests interesting information would be gained from exclusive
final states yields.

The predicted final state-dependent variations of the FWHM widths from the
model used also offer considerable scope for further testing of the
reaction inputs, the shell-model wave functions used, and of the
nucleon-nucleon correlations content of these wave functions.

\section{Summary} \label{sec:summary}

We have considered in detail theoretical expectations for the cross
sections of two-nucleon removal reactions, from \nuc{12}{C} incident
on a carbon target, at beam energies of 250, 1050 and 2100 MeV per
nucleon. The calculated inclusive cross sections for two
like-nucleon ($T=1$) removal are broadly consistent with available
experimental data. For $np$-pair removal the analogous $T=1$,
\nuc{10}{B} final states are also shown to be insensitive to the
shell-model interactions used. The calculated inclusive $np$-pair
removal cross sections at all energies {\em underestimate} the data
by approximately a factor of two. Theoretical calculations of the
widths of the final-state-inclusive residue momentum distributions
on the other hand are consistent with the available experimental
data, including for the $np$-removal channel.

Further measurements, of final-state-exclusive cross sections and
residue momentum distributions, would allow a much more detailed
scrutiny and validation of the direct nature of the reaction,
including the identification of any indirect reaction components.
The calculated $np$-removal cross sections to the $T=0$, \nuc{10}{B}
final states were also shown to have sensitivity to the shell-model
effective interaction used; for example, the ratio of the calculated
cross sections to the \nuc{10}{B} ground $(3^+, T=0)$ and first
$(1^+, T=0)$ excited states.

The overall conclusion from the present analysis is that the
existing data suggest that the $T=0$, $np$-spatial correlations
present in the wave functions used are insufficient. We have shown
that exclusive measurements would offer a means to interrogate these
shell-model inputs, in particular the $np$-channel, $T=0$ wave
functions, and the direct reaction mechanism predictions in
considerable detail.

A similar two-nucleon knockout study can also be performed using 
an \nuc{16}{O} projectile, for which np-pair knockout using 
electromagnetic probes has been measured \cite{MAB06,MAB10}
and that supports the SRC observations for \nuc{12}{C}. In 
this case the \nuc{14}{O} residue is also of interest, given 
that only the $0^+$ ground state is bound, and where the results of 
Refs. \cite{LGH75,GLH75} indicate a correspondingly small cross 
section and a narrow momentum distribution, consistent 
with theoretical expectations. Additionally, possible $2\hbar\omega$ 
components in the \nuc{16}{O} ground state wave function add interest 
to the magnitudes of the absolute cross sections in this case.

\begin{acknowledgments}
The assistance of B.A. Brown and P. Fallon is acknowledged. This
work was supported by the United Kingdom Science and Technology
Facilities Council (STFC) under Grant No. ST/F012012. E.C.S is
grateful for support from the United Kingdom Engineering and
Physical Sciences Research Council (EPSRC) under Grant Number
EP/P503892/1.
\end{acknowledgments}

\end{document}